\documentclass[aps,pra,floatfix,twocolumn,amsmath,amssymb,showpacs]{revtex4-1}

\usepackage{graphicx}
\usepackage{pst-all,pstricks-add}

\newcommand{\unit}[1]{\, \textrm{#1}}

\newcommand{\ket}[1]{\lvert #1 \rangle}           
\newcommand{\bra}[1]{\langle #1 \lvert}           
\newcommand{\innerprod}[2]{\left< #1 \vert #2 \right>}
\newcommand{\commut}[2]{\bigl[ #1, #2 \bigr]}

\newcommand{\eps}{\varepsilon}                 

\newcommand{\rhoi}{\hat{\rho}_\mathrm{0}}
\newcommand{\rhof}{\hat{\rho}_\mathrm{f}}
\newcommand{\rhofconst}{\hat{\rho}_{\mathrm{f} \, \mathrm{const}}}
\newcommand{\rhofbal}{\hat{\rho}_{\mathrm{f} \, \mathrm{bal}}}
\newcommand{\rhotherm}{\hat{\rho}_\mathrm{th}}
\newcommand{\rhodev}{\hat{\rho}_\mathrm{dev}}
\newcommand{\piconst}{\hat{\pi}_\mathrm{const}}
\newcommand{\pibal}{\hat{\pi}_\mathrm{bal}}
\newcommand{\pconst}{p_\mathrm{const}}
\newcommand{\pbal}{p_\mathrm{bal}}

\DeclareMathOperator{\Trace}{Tr}

\begin{document}

\author{David Collins}
\affiliation{Department of Physical and Environmental Sciences, Mesa State College, 1100 North Avenue, Grand Junction, CO 81501, USA}
\email{dacollin@mesastate.edu}
\thanks{Author to whom correspondence should be addressed.}

\title{Discrimination of unitary transformations in the Deutsch-Jozsa algorithm.}

\begin{abstract}
 We describe a general framework for regarding oracle-assisted quantum algorithms as tools for discriminating between unitary transformations. We apply this to the Deutsch-Jozsa problem and derive all possible quantum algorithms which solve the problem with certainty using oracle unitaries in a particular form. We also use this to show that any quantum algorithm that solves the Deutsch-Jozsa problem starting with a quantum system in a particular class of initial, thermal equilibrium-based states of the type encountered in solution state NMR can only succeed with greater probability than a classical algorithm when the problem size exceeds $n \sim 10^5.$
\end{abstract}

\pacs{03.67.Lx}

\maketitle

\section{Introduction}
\label{sec:intro}

Quantum algorithms~\cite{nielsen00} are typically described in terms of the evolution of the state of a quantum system under a prescribed sequence of unitary transformations, followed by the extraction of a problem solution from the outcome of a measurement performed on the quantum system. While the unitary transformations involved are clearly crucial, they are not the primary reference point of the analysis in such a circuit formulation of quantum algorithms; this belongs to the state of the system. However, in the circuit formulation of any particular quantum algorithm, the ingredient which varies from one application of the algorithm to another (i.e.\ the input) is typically a unitary transformation. Therefore, it seems appropriate to focus on the unitaries in a quantum algorithm and to regard the algorithm as a tool for discrimination of unitary transformations. 

This approach has been suggested before~\cite{childs99,vedral07} and subject to analysis in selected scenarios~\cite{bergou05,chefles07}. The purpose of the latter articles was to extend the Deutsch-Jozsa algorithm by investigating the possibility of discriminating amongst a larger class of unitary transformations than that encountered in the original Deutsch-Jozsa algorithm. Our purpose is to further promote  unitary (and state) discrimination as tools for analyzing quantum algorithms. Specifically we consider the Deutsch-Jozsa problem, which is to be solved with the aid of an oracle unitary of a specific form. We ask whether the notions of unitary discrimination can be used to reach the standard quantum algorithm for solving the problem, whether they yield alternative algorithms for solving the problem and what restrictions they impose on quantum algorithms which are implemented on quantum systems initially in noisy mixed states.   

The remainder of this paper is organized as follows. In Sec.~\ref{sec:oraclediscrimination} we describe how oracle-assisted quantum algorithms can be viewed as unitary discrimination tools. This is applied to the Deutsch-Jozsa problem, using a particular oracle unitary,  in two distinct ways in the following sections. In Sec.~\ref{sec:djalg} we use unitary and state discrimination to arrive at the set of all algorithms which solve the Deutsch-Jozsa problem with certainty. In Sec.~\ref{sec:djmixed} we assume a restricted set of possible initial states, which are mixed, for solving the Deutsch-Jozsa problem. We determine a lower bound on the problem size, beneath which a classical algorithm will succeed in solving the Deutsch-Jozsa problem with greater certainty than any quantum algorithm.


\section{Discrimination of quantum operations in oracle algorithms}
\label{sec:oraclediscrimination}

Certain computational problems, such as searching, Simon's problem~\cite{simon94} and the Deutsch-Jozsa problem~\cite{deutsch92}, are oracle-assisted, meaning that they are to be solved using a binary oracle function $f:\{0,1\}^n \mapsto \{0,1\}^m,$ whose form depends on the nature (and the specific instance that is invoked) of the computational problem. The task is to solve the problem with the fewest oracle invocations; the efficiency of the solution is quantified by the number of oracle invocations used. Different instances of a given problem correspond to different oracle functions.

The associated quantum algorithms~\cite{deutsch92,simon94,shor97,grover97} require a well-determined number, depending on $n,$ of qubits prepared in a suitable initial state $\ket{\Psi_\textsf{0}}$. These qubits are made to evolve collectively in way described by a specific sequence of unitary transformations. Ultimately a measurement yields an outcome from which the problem solution can be extracted with high probability. In oracle-based quantum algorithms the oracle is invoked via evolution described by a unitary transformation $\hat{U}_f,$ whose form depends on the problem as well as the particular oracle function, $f$, in use, i.e.\ the particular instance of the problem. In the simplest cases, to which we restrict our consideration, the target of the oracle is a single binary variable, i.e.\ $m=1$. Examples include  the Deutsch-Jozsa~\cite{deutsch92} and Grover's search algorithm~\cite{grover97}. For a given type of problem, there can be different possibilities for the number of qubits required and the structure of the oracle unitary. We consider instances where the number of qubits required is $n.$ In terms of computational basis states, $\ket{x} := \ket{x_n} \otimes \ldots \otimes \ket{x_1}$ with $x_i\in \left\{0,1\right\}$ we assume that the oracle unitary operates via 
\begin{equation}
 \hat{U}_f\ket{x} = \left( -1\right)^{f(x)}\, \ket{x}
 \label{eq:oracleunitary}	
\end{equation}
 and this is extended linearly to all linear combinations of computational basis states. 

\begin{figure}[h]
  \includegraphics[scale=.70]{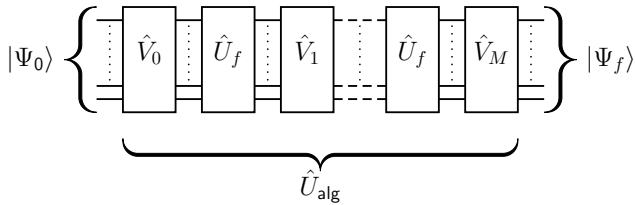}
  \caption{General structure for oracle-based quantum algorithm where the oracle output is a single bit. The oracle is invoked $M$ times. The initial and final unitaries, $\hat{V}_0, \hat{V}_M$ are not generally necessary but are enable initialization and  measurement in the computational basis. 
    \label{fig:generalalg}}
 \end{figure}

Applications of the oracle unitary may be interspersed with other unitaries, $\hat{V}_0, \ldots, \hat{V}_M$ which are \emph{oracle independent}, i.e.\ these remain fixed for all choices of oracle function $f.$ The resulting algorithm has the structure illustrated in Fig.~\ref{fig:generalalg} and the \emph{algorithm unitary} is $\hat{U}_\mathrm{alg} =  \hat{V}_M \hat{U}_f \ldots \hat{U}_f \hat{V}_1 \hat{U}_f \hat{V}_0.$
This is the most general structure for $m=1$ oracle-assisted algorithms since two successive applications of the oracle unitary provide a trivial identity unitary. The final state $\ket{\Psi_f} = \hat{U}_\mathrm{alg} \ket{\Psi_\textsf{0}}$ depends on the particular oracle function~$f$. 

In such scenarios the \emph{input to the algorithm is the oracle function} and not the initial state. The algorithm output then identifies or classifies the input oracle function. For example, in the Deutsch-Jozsa problem the algorithm determines whether the oracle function belongs to the class of constant or balanced functions (these terms are described below). In the algorithm for searching a database with one marked item, located at $s,$ the oracle is defined as $f(x) =0$ if $x\neq s$ and $f(s)=1.$ Determining $s$ is equivalent to identifying which of the possible oracle functions was used. Analogously, the associated quantum algorithms amount to tools for classifying or discriminating between the possible oracle unitaries.

The framework for unitary discrimination requires a set of known unitaries $\left\{ \hat{U}_1, \hat{U}_2, \ldots \right\}$ and associated probabilities $\left\{ p_1, p_2, \ldots \right\}.$ One party chooses one of these unitaries, $\hat{U}_j,$ with probability $p_j$ and another party must determine which unitary was selected by applying the unitary one or more times, together with other quantum operations, to a quantum system. Ultimately a measurement is performed on the quantum system and the choice of unitary is inferred from the measurement outcome. Unitary discrimination can be reduced to a quantum state discrimination problem~\cite{acin01,dariano01,sacchi05,chefles07} by applying the unitary to a standard initial state, most generally described by density operator $\hat{\rho}_\mathrm{0},$ and attempting to discriminate between the possible resulting output states. In this article we consider cases where \emph{the unitary is applied once only.} Thus the possible output states are $\hat{\rho}_{j} = \hat{U}_j \hat{\rho}_\mathrm{0} \hat{U}_j^\dagger$, occurring with probability $p_j,$ for $j=1,2,\ldots.$ The general framework for discriminating between states~\cite{hellstrom76,barnett09} requires a POVM with positive operator elements $\left\{ \hat{\pi}_1, \hat{\pi}_2, \ldots \right\}$ that satisfy $\sum_j \hat{\pi}_j = \hat{I}$ and a rule for associating states with measurement outcomes. In the \emph{minimum error discrimination} scenario, which we consider here, we are required to select one state for each outcome and the ``undecided'' inference of the unambiguous discrimination~\cite{barnett09} scenario is not permitted. Here it is possible to make an incorrect inference and the task is to choose measurement that minimize the probability with which such an error occurs. For unitary discrimination, both the measurement and the initial state must be chosen so as to minimize the error probability.


\section{Application to the Deutsch-Jozsa algorithm on arbitrary initial states}
\label{sec:djalg}

The Deutsch-Jozsa problem consides oracle functions $f: \{0,1\}^n \mapsto \{0,1\}$ that are required to fall into one of two classes: \emph{constant}, meaning that $f$ returns the same value for all possible arguments or \emph{balanced}, meaning that $f$ returns $0$ for exactly half of the arguments and $1$ for the other half. The task is to determine the function class with the  minimum number of oracle invocations. A classical algorithm proceeds by evaluating $f$ at randomly chosen distinct arguments. This will succeed with certainty in all cases after $N/2+1$ oracle invocations~\cite{deutsch92,cleve98}, where $N=2^n$ is the number of possible argument values. A quantum algorithm exists and~\cite{deutsch92,cleve98}, in its modified form~\cite{collins98}, uses an oracle unitary of the form given in Eq.~\eqref{eq:oracleunitary} exactly once to solve the problem with certainty~\cite{deutsch92,cleve98}, giving an exponential speed-up in terms of $n$.

We aim to use the notions of unitary discrimination to arrive at all quantum algorithms which can determine function class with certainty while using $n$ qubits and an oracle of the form of Eq.~\eqref{eq:oracleunitary}. Since each function class is represented by many unitaries, this requires discrimination between two quantum operations. For each given class of functions, the quantum operation is
\begin{equation}
	\hat{\rho}_\mathrm{0} \mapsto \rhof :=\sum_{f\, \textrm{in class}} p_f \hat{U}_f \rhoi \hat{U}_f^\dagger
\end{equation}
where the sum is over all possible functions in the given class and $p_f$ is the probability with which each function could be chosen given that the particular class is chosen. For \emph{constant functions,} Eq.~\eqref{eq:oracleunitary} implies that $\hat{U}_f =\hat{I}$ and thus
\begin{equation}
	\rhoi  \mapsto \rhofconst = \rhoi.
	\label{eq:rhoconst}
\end{equation}
For balanced functions, expanding in the computational basis, $\rhoi= \sum_{x,y=0}^{N-1} \rho_{\mathrm{0}\; xy}\ket{x}\bra{y},$ gives 
\begin{equation}
	\rhoi  \mapsto \rhofbal = \sum_{f\, \textrm{balanced}} p_f 
	           \sum_{x,y=0}^{N-1}
	            \left( -1\right)^{f(x) + f(y)} 
	            \rho_{\mathrm{0}\, xy}
	            \ket{x}\bra{y}.
	\label{eq:rhobalone}
\end{equation}
which gives the density matrix elements  after the balanced function operation $\rho_{\mathrm{f} \, \mathrm{bal}\, xy} := \bra{x} \rhofbal \ket{y}$ as
\begin{equation}
	\rho_{\mathrm{f} \, \mathrm{bal}\, xy} = \sum_{f\, \textrm{balanced}} p_f 
	            \left( -1\right)^{f(x) + f(y)} 
	            \rho_{\mathrm{0}\; xy}.
	\label{eq:rhobaltwo}
\end{equation}
Thus $\rho_{\mathrm{f} \, \mathrm{bal}\, xx} = \rho_{\mathrm{0}\; xx}$ for $x = 0, \ldots, N-1$. For non-diagonal density matrix entries, the summation will be complicated by the possibly different probabilities for each balanced function. We shall assume that the probabilities with which each balanced function is selected are identical. Thus $p_f = 1/B$ where $B$ is the number of balance functions. Enumerating the number of balanced function amounts to counting the number of ways in which $N/2$ of the $N$ possible arguments which will return $0$ can be selected. Thus $B = \binom{N}{N/2}$ and
\begin{equation}
	\rho_{\mathrm{f} \, \mathrm{bal}\, xy} = \frac{1}{B}\; 
	                                         \rho_{\mathrm{0}\; xy} 
	                                         \sum_{f\, \textrm{balanced}} 
	                                         \left( -1\right)^{f(x) + f(y)}.
	\label{eq:rhobalthree}
\end{equation}
The sum $\sum_{f\, \textrm{balanced}} \left( -1\right)^{f(x) + f(y)}$ can be evaluated for given values of $x$ and $y$ by determining for how many balanced functions $f(x) = f(y)$ and for how many $f(x) \neq f(y).$ If $x \neq y,$ this is independent of the choices of $x$ and $y$. This can be established by noting that the collection of all balanced functions can be listed by assigning $0$ to $N/2$ of the $N$ possible argument ``slots'' and $1$ to the remaining slots. The argument values, $x=0,1,2,\ldots N-1$ in this process merely serve as labels and interchanging two of them will not affect the sum. Thus it suffices to compute this for $x=0$ and $x=1.$ The four possibilities are tabulated in Table~\ref{tab:balanced}.

\begin{table}[h]
 \begin{ruledtabular}
 \begin{tabular}{cccc}
   $f(x=0)$ & $f(y=1)$ &  $\left( -1\right)^{f(x) + f(y)}$ & Number of Instances \\
     \hline
     $0$ & $0$ & $1$ & $\binom{N-2}{N/2}$\\ 
     $0$ & $1$ & $-1$ & $\binom{N-2}{N/2 -1}$\\ 
     $1$ & $0$ & $-1$ & $\binom{N-2}{N/2 -1}$\\ 
     $1$ & $1$ & $1$ & $\binom{N-2}{N/2}$\\ 
 \end{tabular}
 \end{ruledtabular}
 \caption{Four possibilities for $\left( -1\right)^{f(x) + f(y)}$ for $x=0$ and $y=1.$ The two leftmost columns provide the possible combinations of values returned by $f.$ The last column lists the number of times that each possibility occurs. For example,  the balanced function for which $f(0) = f(1) = 0$ must return $1$ in $N/2$ of the remaining $N-2$ arguments. The number of ways in which this arises is $\binom{N-2}{N/2}.$ Similar arguments apply to the other cases.
         \label{tab:balanced}
         }
\end{table}

Thus, if $x\neq y$ then 
\begin{align}
   \sum_{f\, \textrm{balanced}} \left( -1\right)^{f(x) + f(y)} & = 2
                                                                 \left[
                                                                   \binom{N-2}{N/2} - \binom{N-2}{N/2-1}
                                                                 \right] \nonumber \\
                                                               & = -\binom{N-2}{N/2}\;
                                                                    \frac{2}{N/2-1} \nonumber \\
                                                               & = - \frac{B}{N-1}
	\label{eq:faddition}
\end{align}
where $B$ is the number of balanced functions and the last two lines follow from algebraic manipulations of combinatorials. Eqs.~\eqref{eq:rhobalthree} and \eqref{eq:faddition} imply that, if $x\neq y$,  
\begin{equation}
	\rho_{\mathrm{f} \, \mathrm{bal}\, xy} = - \frac{1}{N-1}\; 
	                                         \rho_{\mathrm{0}\; xy}.
	\label{eq:rhobalfour}
\end{equation}
The cases of all values of $x$ and $y$ are then summarized as 
\begin{equation}
	\rhofbal = \frac{1}{N-1}\,
	           \left( 
	             -\rhoi + N \sum_{x=0}^{N-1} \hat{P}_x \rhoi \hat{P}_x
	           \right)
	\label{eq:rhobalfive}
\end{equation}
where $\hat{P}_x := \ket{x}\bra{x}.$

Minimum error discrimination  between the two density operators $\rhofconst$ and $\rhofbal$ requires a POVM with two outcomes and two positive operator elements $\piconst$ and $\pibal,$ where $\piconst + \pibal = \hat{I}.$ The probability with which an incorrect inference is made is 
\begin{equation}
 p_\mathrm{error} = \pconst \Trace{\left[\rhofbal\piconst\right]}
                  + \pbal \Trace{\left[\rhofconst\pibal\right]}
 \label{eq:errordensity}
\end{equation}
where $\pconst$  ($\pbal$) is the probability of selecting a function from the constant (balanced) class. A standard derivation~\cite{sacchi05} gives
\begin{equation}
	p_\mathrm{error} = \frac{1}{2}\; 
	                   \left( 1 - \lVert \pconst \rhofconst - \pbal \rhofbal \rVert \right)
	\label{eq:errorprob}
\end{equation}
where the trace norm satisfies 
\begin{equation}
	\lVert \hat{A} \rVert := \Trace{\left[ \sqrt{\hat{A}^\dagger \hat{A}}\; \right]} = \sum_i \sigma_i(\hat{A})
\end{equation}
with $\left\{ \sigma_i(A) \right\}$ being the singular values of $\hat{A}.$ The trace norm is clearly invariant under unitary transformations in the sense that, if $\hat{V}$ is any unitary then $\lVert \hat{A} \rVert = \lVert \hat{V} \hat{A}\hat{V}^\dagger \rVert.$ Although Eqs.~\eqref{eq:errordensity} and~\eqref{eq:errorprob} are equivalent, they have distinct uses in terms of determining conditions under which the algorithm will succeed. As we shall show, Eq.~\eqref{eq:errorprob} yields the optimal initial state and Eq.~\eqref{eq:errordensity}, the optimal measurement for success.

Eq.~\eqref{eq:errorprob} implies that the quantum algorithm for solving the Deutsch-Jozsa problem will succeed with certainty when $\lVert \pconst \rhofconst - \pbal \rhofbal \rVert=1.$ Note that $\lVert \pconst \rhofconst \rVert =\pconst$ and $\lVert \pbal \rhofbal \rVert =\pbal$, giving $\lVert \pconst \rhofconst \rVert + \lVert \pbal \rhofbal \rVert = \pbal + \pconst =1.$ Thus the quantum algorithm will succeed with certainty if and only if
\begin{equation}
	\lVert \pconst \rhofconst - \pbal \rhofbal \rVert = \lVert \pconst \rhofconst \rVert 
	                                                  + \lVert \pbal \rhofbal \rVert.
\end{equation}
In this context, an important general result~\cite{qiu08} is that if $\hat{A}$ and $\hat{B}$ are positive semidefinite operators then $\lVert \hat{A} - \hat{B}\rVert = \lVert \hat{A} \rVert + \lVert \hat{B} \rVert$ if and only if $\hat{A}$ and $\hat{B}$ have orthogonal support. The support of a positive semidefinite operator is the subspace spanned by its eigenstates which correspond to non-zero eigenvalues. Thus the quantum algorithm will succeed with certainty if and only if the operators $\rhofconst$ and $\rhofbal$ have orthogonal support, or equivalently
\begin{equation}
	\rhofconst\rhofbal = \rhofbal\rhofconst =0.
	\label{eq:orthogrho}
\end{equation}

Defining $\hat{\Lambda} : = \sum_{x=0}^{N-1} \hat{P}_x \rhoi \hat{P}_x$, which is easily shown to be a positive operator, and using Eqs~\eqref{eq:rhoconst}, \eqref{eq:rhobalfive} and~\eqref{eq:orthogrho} gives
\begin{align}
   \commut{\rhoi}{\hat{\Lambda}}& = 0 \quad \textrm{and} \label{eq:condone}\\
   N \hat{\Lambda} \rhoi  & = \rhoi^2.  \label{eq:condtwo}
\end{align}
Eq~\eqref{eq:condone} implies that $\rhoi$ and $\hat{\Lambda}$ can be simultaneously diagonalized. Denote the associated basis of eigenstates by $\{ \ket{\phi_j} \; | \; j = 1,\ldots N \}.$ Thus
\begin{equation}
	\rhoi = \sum_{j=1}^L r_j \ket{\phi_j}\bra{\phi_j}
\end{equation}
where $L>0$ is the number of non-zero eigenvalues of $\rhoi$ and $0< r_j \leqslant 1$ satisfy $\sum_{j=1}^L r_j = 1.$ Likewise
\begin{equation}
	\hat{\Lambda} = \sum_{j=1}^N \lambda_j \ket{\phi_j}\bra{\phi_j}
	\label{eq:lambdadiag}
\end{equation}
where $\lambda_j \geqslant 0$. Eq.~\eqref{eq:condtwo} implies that $N \lambda_j r_j = r_j^2$ for $j=1, \ldots, L$ and this gives
\begin{equation}
	\lambda_j = \frac{r_j}{N}
	\label{eq:lambdar}
\end{equation}
for $j=1, \ldots, L$. Additionally, 
\begin{align}
	\hat{\Lambda} & = \sum_{x=0}^{N-1} 
	                  \hat{P}_x 
	                  \sum_{j=1}^L r_j \ket{\phi_j}\bra{\phi_j} 
	                  \hat{P}_x \\
	              & = \sum_{x=0}^{N-1} 
	                  \hat{P}_x 
	                  \sum_{j=1}^L
	                  r_j \lvert \phi_j(x) \rvert^2
\end{align}
where $\phi_j(x):= \innerprod{x}{\phi_j}.$ Thus Eq.~\eqref{eq:lambdadiag} gives
\begin{equation}
	\lambda_k = \sum_{x=0}^{N-1} 
	                  \sum_{j=1}^L
	                  r_j 
	                  \lvert \phi_k(x) \rvert^2
	                  \lvert \phi_j(x) \rvert^2
\end{equation}
and, when combined with Eq.~\eqref{eq:lambdar}, 
\begin{equation}
	\frac{r_k}{N} = \sum_{x=0}^{N-1}
	                  \sum_{j=1}^L
	                  r_j 
	                  \lvert \phi_k(x) \rvert^2
	                  \lvert \phi_j(x) \rvert^2
\end{equation}
for $k\leqslant L.$ The fact that $L>0$ implies that $r_1 \neq 0.$ Thus
\begin{equation}
	\frac{r_1}{N} =   r_1\sum_{x=0}^{N-1} 
	                  \lvert \phi_1(x) \rvert^4
	                 +\sum_{x=0}^{N-1} 
	                  \sum_{j=2}^L
	                  r_j 
	                  \lvert \phi_k(x) \rvert^2
	                  \lvert \phi_j(x) \rvert^2
	 \label{eq:firsteigenval}
\end{equation}
where the second term on the right is defined to mean $0$ when $L=1.$
The second term on the right of Eq.~\eqref{eq:firsteigenval} is non-negative. The first term contains $\sum_{x=0}^{N-1} \lvert \phi_1(x) \rvert^4$ which is subject to the constraint that $\sum_{x=0}^{N-1} \lvert \phi_1(x) \rvert^2 = 1.$ A Lagrange multiplier approach shows that $\sum_{x=0}^{N-1} \lvert \phi_1(x) \rvert^4 \geqslant 1/N$, provided that $0 \leqslant \lvert \phi_1(x) \rvert^2 \leqslant 1,$ which is always satisfied. This minimum is attained when $\lvert \phi_1(x) \rvert^2 = 1/N$ (or equivalently $\phi_1(x) = e^{i\theta_x}/\sqrt{N}$ where $\theta_x$ is real) for  $x = 0, \ldots N-1$. Thus the second term on the right of Eq.~\eqref{eq:firsteigenval} is identically zero and $L=1.$ Thus the quantum algorithm will succeed with certainty if and only if the input is a pure state, i.e.\ (dropping the subscript and changing notation to be consistent with Fig.~\ref{fig:generalalg})
\begin{equation}
	\rhoi = \ket{\Psi_0}\bra{\Psi_0}.
\end{equation}
The pure state must be such that it produces a minimum for the first term on the right of Eq.~\eqref{eq:firsteigenval} and thus 
\begin{equation}
	\ket{\Psi_0} = \frac{1}{\sqrt{N}}
	       \sum_{x=0}^{N-1}
	       e^{i\theta_x}
	       \ket{x}.
	\label{eq:optimalstate}
\end{equation}

Eq.~\eqref{eq:errordensity} and the positivity of the POVM elements and density operators requires that for the algorithm to succeed with certainty the measurement operators must satisfy $\Trace{\left[\rhofbal\piconst\right]} = \Trace{\left[\rhofconst\pibal\right]}=0.$ But $\piconst = \hat{I} - \pibal$ implies implies that this requirement is equivalent to $\Trace{\left[\rhofconst\piconst\right]} = \Trace{\left[\rhofbal\pibal\right]} = 1.$ The fact that the density operators and the measurement operators are positive semidefinite and that their eigenvalues fall in the range $\left[ 0,1\right]$ then implies that $\piconst$ must be the projector onto the support of $\rhofconst.$ The associated POVM elements are
\begin{subequations}
\begin{align}
	\piconst & = \ket{\Psi_0}\bra{\Psi_0} \quad \textrm{and}\\
	\pibal & = \hat{I} - \ket{\Psi_0}\bra{\Psi_0}.
\end{align}
	\label{eq:optimalPOVM}
\end{subequations}

Eqs.~\eqref{eq:optimalstate} and~\eqref{eq:optimalPOVM} give the initial states and measurements for a general quantum algorithm which solves the Deutsch-Jozsa problem using a single invocation of an oracle unitary of Eq.~\eqref{eq:oracleunitary} and no ancillary qubits. The standard algorithm, for which $\ket{\Psi_0} = \frac{1}{\sqrt{N}} \sum_{x=0}^{N-1} \ket{x},$ is one example of this~\cite{cleve98,collins98}. Also the general algorithm will successfully identify the function class regardless of the probabilities with which the various admissible functions are selected. This can be verified by applying the algorithm unitary transformations for the various admissible functions to the initial state of Eq.~\eqref{eq:optimalstate}, computing the final state and using this to determine the POVM operators of Eq.~\eqref{eq:optimalPOVM} to determine probabilities of the two possible outcomes. 


\section{Application to the Deutsch-Jozsa algorithm on mixed initial states}
\label{sec:djmixed}

In some proposed implementations of quantum computing, such as solution state nuclear magnetic resonance (NMR)~\cite{chuang98a,cory98,cory97,gershenfeld97,marx00,vdsypen01,vdsypen00,vdsypen04,negrevergne05,negrevergne06}, the initial state of the quantum system is mixed and therefore the problem cannot be solved with certainty by using the scheme of Fig.~\ref{fig:generalalg}. The notion of minimum error discrimination between quantum operations can be applied to bound the success probability of any algorithm on such mixed input states. For a general mixed state, Eqs.~\eqref{eq:rhoconst}, \eqref{eq:rhobalfive} and \eqref{eq:errorprob} imply  
\begin{equation}
  p_\mathrm{error} = \frac{1}{2}\; 
	                   \left( 1 - \left\lVert 
	                               \left( 
	                                 \pconst + \frac{\pbal}{N-1} 
	                               \right) \rhoi - \frac{\pbal N}{N-1} \hat{\Lambda} \right\rVert \right).
\end{equation}
We consider the special case where $\pconst = \pbal = 1/2$ and thus
\begin{equation}
  p_\mathrm{error} = \frac{1}{2}\; 
	                   \left[ 1 - \frac{N}{2(N-1)}\; 
	                              \left\lVert
	                                \rhoi - \hat{\Lambda} 
	                              \right\rVert 
	                   \right].
\end{equation}
Attempting to apply the triangle inequality to the trace norm and using the facts that $\left\lVert \rhoi \right\rVert = \left\lVert \hat{\Lambda} \right\rVert =1$ results in an inequality which is always satisfied.

However, in some circumstances a similar approach can yield meaningful bounds. Specifically we consider situations in which an ensemble of quantum systems is first allowed to reach thermal equilibrium followed by application of a preparatory, oracle-independent unitary. The resulting state constitutes the initial state for the system. We shall assume that the system consists of $n$ qubits and has a Hamiltonian 
\begin{equation}
	\hat{H} = \sum_{i=1}^n \frac{\hbar \omega_i }{2} \hat{\sigma}_z^{(i)} 
	        + \sum_{i<j}^n \frac{\hbar \pi J_{ij}}{2} \hat{\sigma}_z^{(i)} \otimes \hat{\sigma}_z^{(j)} 
\end{equation}
where $\omega_i$ is the precession frequency of the $i^\mathrm{th}$ qubit and $J_{ij}$ is the coupling between the $i^\mathrm{th}$ and $j^\mathrm{th}$ qubits (this is typical for solution state NMR~\cite{cory98,chuang98a}). The thermal equilibrium density operator is  $\rhotherm =e^{-\beta \hat{H}}/Z$ where $\beta = 1/kT$ and $Z = \Trace{\left[ e^{-\beta \hat{H}} \right]}.$ We shall assume that $J_{ij} \ll \omega_k$ and we shall ignore coupling terms in the density operator which is constructed from
\begin{equation}
	e^{-\beta \hat{H}} \approx e^{-\sum_{i=1}^n \alpha_i\hat{\sigma}_z^{(i)}  }
\end{equation}
where $\alpha_i:= \hbar \omega_i /2kT.$ Our final assumption is that $\alpha_i \ll 1$ (in typical solution state NMR scenarios, $\omega_i \sim 500\unit{MHz}$ and $T \sim 300\unit{K}$ so that $\alpha_i \sim 10^{-5}$). Thus
\begin{equation}
	\rhotherm \approx \frac{1}{N} \hat{I} - \rhodev.
\end{equation}
where the deviation density operator is $\rhodev := \sum_{i=1}^n \alpha_i \hat{\sigma}_z^{(i)}/N.$ Note that $\Trace{\rhodev}=0.$ Given the thermal equilibrium state as an initial state of the system, there is a range of possible quantum operations that can be applied prior to the first invocation of the oracle. These could contain non-unitary operations; examples include pseudopure state preparation~\cite{chuang98a,cory98} or algorithmic cooling~\cite{boykin02, *schulman05}. Our aim is to focus on scenarios in which such non-unitary operations are avoided and we only consider scenarios where \emph{operations applied prior to the first oracle invocation are unitary without involving any ancillary systems.} 

Here the most general initial state of the $n$ qubit system is represented by $\rhoi = \hat{V} \rhotherm \hat{V}^\dagger$ where $\hat{V}$ is the preparatory unitary operation. Then 
\begin{equation}
	\rhoi = \frac{1}{N} \hat{I} - \rhodev^\prime
\end{equation}
where $\rhodev^\prime = \hat{V}\rhodev \hat{V}^\dagger$ and  
\begin{equation}
	\hat{\Lambda} = \frac{1}{N} \hat{I} - \sum_{x=0}^{N-1} \hat{P}_x \rhodev^\prime \hat{P}_x.
\end{equation}
Define $\hat{\Lambda}_\mathrm{dev}^\prime:= \sum_{x=0}^{N-1} \hat{P}_x \rhodev^\prime \hat{P}_x.$
Thus
\begin{equation}
  p_\mathrm{error} = \frac{1}{2}\; 
	                   \left[ 1 
	                         -\frac{N}{2(N-1)}\;
	                          \left\lVert
	                           \rhodev^\prime 
	                           - 
	                           \hat{\Lambda}_\mathrm{dev}^\prime
	                          \right\rVert 
	                   \right].
\end{equation}
The triangle inequality yields
\begin{equation}
  p_\mathrm{error} \geqslant \frac{1}{2}\; 
	                   \left[ 1 
	                         -\frac{N}{2(N-1)}\;
	                          \left(
	                          \left\lVert
	                           \rhodev^\prime
	                          \right\rVert 
	                           +
	                          \left\lVert 
	                           \hat{\Lambda}_\mathrm{dev}^\prime
	                          \right\rVert 
	                          \right)
	                   \right].
\end{equation}
The diagonal nature of $\hat{\Lambda}_\mathrm{dev}^\prime$ gives
\begin{equation}
	\left\lVert \hat{\Lambda}_\mathrm{dev}^\prime  \right\rVert = \sum_{x=0}^{N-1} 
	                                           \left\lvert \bra{x}\rhodev^\prime \ket{x} \right\rvert.
\end{equation}
Denote the eigenstates and eigenvalues of $\rhodev^\prime$ by $\left\{ \ket{\chi_i} \; | \; i=1,\ldots,N \right\}$ and $\left\{ c_i \; | \; i=1,\ldots,N \right\}$ respectively. Then
\begin{align}
	\left\lVert \hat{\Lambda}_\mathrm{dev}^\prime  \right\rVert
	& = 
	\sum_{x=0}^{N-1} 
	\sum_{i=0}^N
	\left\lvert c_i \right\rvert
	\left\lvert
	  \innerprod{x}{\chi_i}
	\right\rvert^2 \nonumber \\
  & = 
  \sum_{i=0}^N
	\left\lvert c_i \right\rvert
	\sum_{x=0}^{N-1} 
  \left\lvert
	  \innerprod{x}{\chi_i}
	\right\rvert^2 \nonumber \\
	& =  
	\sum_{i=0}^N
	\lvert c_i \rvert
   = \left\lVert \rhodev^\prime  \right\rVert.
\end{align}

Thus
\begin{equation}
	p_\mathrm{error} \geqslant \frac{1}{2}\; 
	                   \left[ 1 
	                         -\frac{N}{(N-1)}\;
	                          \left\lVert
	                           \rhodev^\prime
	                          \right\rVert 
	                   \right].
\end{equation}
The unitary invariance of the trace norm implies that $\left\lVert \rhodev^\prime \right\rVert = \left\lVert \rhodev \right\rVert.$ The singular values of $\rhodev$ are $\left\{ \lvert \alpha_1 + \alpha_2 + \ldots \alpha_n\rvert/N, \right.$ $\left.  \lvert -\alpha_1 + \alpha_2 + \ldots \alpha_n\rvert/N, \lvert \alpha_1 - \alpha_2 + \ldots \alpha_n\rvert/N, \ldots \right\}.$ Without loss of generality assume that $\alpha_1\geqslant \alpha_2 \geqslant \alpha_3, \ldots \geqslant 0.$ Then each singular value is bounded from below by $n \alpha_1/N$ and
\begin{equation}
	\left\lVert \rhodev^\prime  \right\rVert  \leqslant n \alpha_1.
\end{equation}

Thus
\begin{equation}
	p_\mathrm{error} \geqslant \frac{1}{2}\; 
	                   \left( 1 
	                         -\frac{Nn \alpha_1}{N-1}\;
	                   \right).
	\label{eq:errorthermal}
\end{equation}
Defining $\eps:= Nn\alpha_1 /(N-1)$ gives $p_\mathrm{error} \geqslant (1-\eps)/2.$ This gives the failure probability for an application of the algorithm to a single ensemble member and it indicates a non-deterministic output. The algorithm may be run repeatedly using the same unitary on a large number of independent quantum systems, all initially described by the same density operator and this will ultimately increase the chances of successfully identifying the oracle class. However, this must be compared to a classical probabilistic algorithm. Such an analysis has been done in the case of a pseudopure-state input~\cite{anderson05}. Here the quantum algorithm is reconfigured, using a single ancillary qubit to which the function class is written whenever a correct pure state input is used. For mixed input states, it is possible that an erroneous function class may be written to this ancillary qubit. Over an entire ensemble, the function class is inferred by effectively taking a ``majority-vote'' of computational basis measurement outcomes on the ancillary qubits for individual ensemble members. The probability of successful inference only depends on the ensemble size and the probabilities with which each of the two possible measurement outcomes occur; the initial state merely determines these probabilities in terms of $\eps$. This is to be compared with a classical probabilistic algorithm. The results are that, for large ensemble size (i.e.\ in the limit as this becomes infinite) the quantum algorithm succeeds with larger probability than the classical probabilistic algorithm if $\eps > \sqrt{3/4}.$  For the quantum algorithm to succeed with greater probability than the classical algorithm, this implies
\begin{equation}
	\alpha_1 > \sqrt{3/4}\; \frac{N-1}{N}\; \frac{1}{n}
\end{equation}
or, alternatively
\begin{equation}
	n > \sqrt{3/4}\; \frac{N-1}{N}\; \frac{1}{\alpha_1}.
\end{equation}
In typical solution-state NMR situations $\alpha_1 \sim 10^{-5}$ and thus  
\begin{equation}
	n > \sqrt{3/4}\; \frac{N-1}{N}\; 10^5.
\end{equation}
Thus with current solution state NMR technology and an implementation of the algorithm which starts with the thermal equilibrium state, uses no ancillary qubits and only uses the oracle unitary of Eq.~\eqref{eq:oracleunitary} plus any other unitaries, the minimum number of qubits required for the quantum algorithm to succeed with greater probability than any classical algorithm is approximately $10^5$. Current implementations of the Deutsch-Jozsa algorithm have not even exceeded $n=10.$\\

\section{Conclusion}

In conclusion we have demonstrated the usefulness of the notion that quantum algorithms are tools for discriminating between oracle unitary transformations. For the Deutsch-Jozsa problem, applying techniques associated with unitary discrimination results in the complete set of quantum algorithms which can solve the problem with certainty; these have a structure similar to that of the standard version of the algorithm. We have also explored the issue of solving the Deutsch-Jozsa problem when the initial state of the quantum system is selected from a restricted set. For the case of a thermal equilibrium state followed by an arbitrary oracle-independent unitary, this yields a lower bound on the number of function arguments such that the quantum algorithm succeeds with greater probability than any classical algorithm. This is several orders of magnitude larger than that which has been implemented experimentally to date.

We should point out that our analysis assumed certain constraints. First we assumed the oracle unitary of Eq.~\eqref{eq:oracleunitary}, which operates on $n$ qubits. However, some versions of the algorithm require $n$ argument qubits plus one ancilla qubit and use an oracle unitary defined by $\hat{U}_f\ket{x}\ket{y} := \ket{x}\ket{y\oplus f(x)}$ (the rightmost system in this notation is the single ancilla qubit, the leftmost the $n$ argument qubits)~\cite{deutsch92}. The unitary considered in our analysis can be reached from this by fixing the ancilla qubit in a special state, converting the problem to one of phase estimation~\cite{cleve98}. Whether this yields a larger set of quantum algorithms which solves the problem with certainty or reduces the lower bound in the thermal equilibrium state scenario is open to investigation. Second, it is known that using ancilla qubits entangled with ``system'' qubits enhances the possibility of successful discrimination between unitaries~\cite{dariano01} and it is conceivable that this could also yield a larger set of quantum algorithms and improved lower bounds in the thermal equilibrium case. Finally, in the thermal equilibrium scenario, relaxation to the thermal equilibrium state could be followed by a non-unitary quantum operation~\cite{fahmy08} and we have not assessed bounds in any such cases.

\acknowledgments

The author would like to thank Michael Frey for many useful discussions and is grateful for support from the Mesa State Faculty Professional Development Fund.


%

\end{document}